\documentclass[%
 reprint,
groupedaddress,
showpacs,%preprintnumbers,
 amsmath,amssymb,
 aps,
 prd,
 longbibliography,
 lengthcheck,%
]{revtex4-1}

\usepackage{graphicx}% Include figure files
\usepackage{dcolumn}% Align table columns on decimal point
\usepackage{bm}% bold math
\usepackage{float}

\begin{document}

\title{Highly relativistic spin-gravity coupling for fermions}

\author{Roman Plyatsko and Mykola Fenyk}

\affiliation{Pidstryhach Institute for Applied Problems in
Mechanics and Mathematics\\ Ukrainian National Academy of Sciences, 3-b Naukova Street,\\
Lviv, 79060, Ukraine}

\date{\today}

\begin{abstract}
Descriptions of  highly relativistic fermions in a gravitational field in the classical (nonquantum) and quantum approaches are discussed. The results following from the Mathisson-Papapetrou equations for a fast spinning particle in Schwarzschild's and Kerr's background are considered. Numerical estimates for electron, proton and neutrino in the gravitational field of black holes are presented.The general relativistic Dirac equation is analyzed from the point of view it is using for the adequate description of highly relativistic fermions in a gravitational field, in the linear and nonlinear spin approximation. It is necessary to have some corrected Dirac equation for a highly relativistic fermion with strong spin-gravity coupling.

\end{abstract}

%\keywords{}

\maketitle

\section{ Introduction}

One can read the key words {\it spin-gravity coupling} and {\it spin-gravity interaction} one can read in many papers on general relativity, whereas the words {\it highly relativistic} (or {\it ultrarelativistic) spin-gravity coupling} and {\it highly relativistic spin-gravity interaction} are very rare in the corresponding literature. Does it means that there is not essential difference in the reaction of a spinning particle on the gravitational field when its velocity is: 1) much less than the velocity of light and 2) very close to this velocity? The main purpose of this paper is to answer this question.

In the context of this subject we recall  paper \cite{Kobzar} where the
gravitational interaction of spin-1/2 particles was considered in the linear approximation. An important result of this paper is that the gravitational interaction of two fermions increases with increasing energy like $E^2$. Thus, for protons, when $E_p$ is of order $10^{19}m_p$  (here $m_p$ is the proton mass), the gravitational interaction becomes as strong as the electromagnetic one, by comparison of the corresponding amplitudes of scattering \cite{Kobzar}.

Another direction of fermion-gravity coupling investigations is based on the general relativistic Dirac equation which is known since 1929 \cite{Fock}.
The correspondent solutions of this equation in  Schwarzschild's and Kerr's
background are reflected in \cite{Chandra}. Different aspects of the covariant Dirac equation and its solutions are presented in recent papers \cite{Obukh}.

There is a possibility to study the spin-gravity coupling in the classical (nonquantum) approach by the Mathisson-Papapetrou (MP) equations \cite{Mathis, Papa} which describe motions of a spinning test body (particle) in a gravitational field according to general relativity. It is important that in a certain sense the MP equations follow from the general relativistic Dirac equation as a classical approximation \cite{Wong} (see also some papers from \cite{Obukh}). Thus, just the MP equations can be used to investigate some features of spin-gravity coupling for fermions, when their quantum properties are not important.

We stress that the possible role of the fermion's interaction with gravity is often discussed in high energy physics and cosmology. In particular, in the focus of many investigations, both theoretical and experimental,  high energy neutrinos in strong gravitational fields are studied. Here we point out only the most recent publications \cite{Kostel, Bonder} which are especially interesting in the context of our paper.

The paper is organized as follows. In Sec. II the basic information on the MP equations is presented. The concrete relationships which characterize the spin-gravity coupling for a highly relativistic spinning particle in Schwarzschild's field are considered in Sec. III. The situation when at the highly relativistic orbital  velocity some restriction arises on using the supplementary condition for the MP equations in the Schwarzschild background is analyzed in Sec. IV. Section V is devoted to the results following from the MP on possible highly relativistic orbits of a spinning particle both in linear and nonlinear spin approximation in Schwarzschild's and Kerr's background. In Sec. VI we draw attention to the specificity of the description of highly relativistic fermions in a gravitational field by the general relativistic Dirac equation. We conclude in Sec. VII.

\section{Mathisson-Papapetrou equations}

 These equations can be written as  \cite{Mathis, Papa}
\begin{equation}\label{1}
\frac D {ds} \left(mu^\lambda + u_\mu\frac {DS^{\lambda\mu}}
{ds}\right)= -\frac {1} {2} u^\pi S^{\rho\sigma}
R^{\lambda}_{~\pi\rho\sigma},
\end{equation}
\begin{equation}\label{2}
\frac {DS^{\mu\nu}} {ds} + u^\mu u_\sigma \frac {DS^{\nu\sigma}}
{ds} - u^\nu u_\sigma \frac {DS^{\mu\sigma}} {ds} = 0,
\end{equation}
where $u^\lambda\equiv dx^\lambda/ds$ is the particle's 4-velocity,
$S^{\mu\nu}$ is the tensor of spin, $m$ and $D/ds$ are,
respectively, the mass and the covariant derivative along $u^\lambda$ and $R^{\lambda}_{~\pi\rho\sigma}$ is
the Riemann curvature tensor (units $c=G=1$ are used). Here, and in
the following, Latin indices run 1, 2, 3 and Greek indices 1, 2, 3,
4; the signature of the metric (--,--,--,+) is chosen. After \cite{Mathis, Papa} Eqs. (\ref{1}) and  (\ref{2}) were obtained in many papers by different approaches \cite{Tul, Dixon}.

It is necessary to add a supplementary condition to Eqs. (\ref{1}) and  (\ref{2}) in order to choose an appropriate position of the particle's center of mass. Most often the conditions  \cite{Mathis, Pirani}
\begin{equation}\label{3}
S^{\lambda\nu} u_\nu = 0
\end{equation}
and  \cite{Tul}
\begin{equation}\label{4}
S^{\lambda\nu} P_\nu = 0
\end{equation}
are used, where
\begin{equation}\label{5}
P^\nu = mu^\nu + u_\lambda\frac {DS^{\nu\lambda}}{ds}
\end{equation}
is the particle 4-momentum. Both at conditions (\ref{3}) and (\ref{4}), the constant of motion of the MP
equations is
\begin{equation}\label{6} S_0^2=\frac12
S_{\mu\nu}S^{\mu\nu},
\end{equation}
where $|S_0|$ is the absolute value of spin.

Often instead of Eqs. (\ref{1}) and  (\ref{2}) their linear spin approximation is considered:
\begin{equation}\label{7}
m\frac D {ds} u^\lambda = -\frac {1} {2} u^\pi S^{\rho\sigma}
R^{\lambda}_{~\pi\rho\sigma},
\end{equation}
\begin{equation}\label{8}
\frac {DS^{\mu\nu}} {ds}  = 0.
\end{equation}
In this approximation condition (\ref{4}) coincides with (\ref{3}) and $m$ is the constant of motion [for the exact MP equations $m$ is the constant of motion only at condition (\ref{3}), and under condition (\ref{4}) the constant of motion is the value $\mu^2 \equiv P_\lambda P^\lambda$].
We stress that just Eqs. (\ref{7}) and (\ref{8}) follow from the Dirac equation by some procedure of reduction with the Dirac matrices \cite{Obukh}.

To better understand the nature of the spin-gravity coupling it is useful to consider the MP equations as written in the comoving tetrad representation
\cite{Pl98}. In particular, it follows from (\ref{7}) that
\begin{equation}
\label{9}
\gamma_{(i)(4)(4)} = -\frac{s_{(1)}}{m} R_{(1)(4)(2)(3)},
\end{equation}
where $\gamma_{(\alpha)(\beta)(\delta)}$ are Ricci's coefficients of rotation,
$s_{(1)}$ and $R_{(1)(4)(2)(3)}$ are, respectively, the local tetrad components of the particle's spin 4-vector and the local components of the Riemann tensor; here the first local coordinate axis is orientated along the spin, and indices of the tetrad are placed in parentheses. We use the definition of Ricci's coefficients as $\gamma_{(\alpha)(\beta)(\delta)}=\lambda_{(\alpha)\mu;\nu}\lambda_{(\beta)}^{\mu}\lambda_{(\delta)}^{\nu}$, where $\lambda_{(\alpha)\mu}$ are the orthogonal tetrads and " ; " notes the covariant derivative.  (By definition,
$s_\lambda=\frac12 \sqrt{-g}\varepsilon_{\lambda\mu\nu\sigma}u^\mu
S^{\nu\sigma}$, where $s_\lambda$ is 4-vector of spin in the global coordinates; $g$ and $\varepsilon_{\lambda\mu\nu\sigma}$ are the determinant of the metric tensor and the symbol Levi-Civita, respectively.)

It is important that Ricci's coefficients of rotation
$\gamma_{(i)(4)(4)}$ have the direct physical meaning as the components
$a_{(i)}$ of the 3-accelerate of a spinning test particle relative to geodesic free fall as measured by the observer comoving with this particle. Therefore, by Eq. (\ref{9}) we have
\begin{equation}
\label{10}
a_{(i)} = -\frac{s_{(1)}}{m} R_{(1)(4)(2)(3)}.
\end{equation}

It follows from  Eq. (\ref{8}) that $\gamma_{(i)(k)(4)}=0$, i.e., the known condition for the Fermi transport in terms of the Ricci coefficients of rotation.

\section{Spin-gravity coupling at high velocity: an example}

Let us estimate the action of the spin-gravity coupling on a spinning particle moving with some velocity in Schwarzschild's background. For this purpose we take into account the expressions for different components of the gravitational field of a moving Schwarzschild source. Namely, following \cite{Thorne:1985} we  use the general definition of the gravitoelectric
 $E_{(k)}^{(i)}$ and gravitomagnetic   $B_{(k)}^{(i)}$ components of the gravitational field as
\begin{equation}\label{11}
E_{(k)}^{(i)}=R^{(i)(4)}_{}{}{}{}{}{}_{(k)(4)},
\end{equation}
\begin{equation}\label{12}
B_{(k)}^{(i)}=-\frac12 R^{(i)(4)}_{}{}{}{}{}{}_{(m)(n)}
\varepsilon^{(m)(n)}_{}{}{}{}{}{}_{(k)}.
\end{equation}
(Note that in the context of the analysis of the MP equations, another definition of the gravitoelectric and gravitomagnetic fields as the corresponding 3-vector values with world indices is effectively used in \cite{Obukh}.)

According to (\ref{10}) and (\ref{12}) from the point of view of the observer which is comoving with the spinning particle we have
\begin{equation}\label{13}
a_{(i)}=-\frac{s_{(1)}}{m}B^{(1)}_{(i)}.
\end{equation}

In the concrete case of Schwarzschild's mass, when the first space local vector is orthogonal to the plane that is determined by the instantaneous direction of the observer motion relative to the mass and the radial direction, and the second vector is directed along the direction of the motion of the observer, there are the expressions for the nonzero components of the gravitomagnetic field \cite{Pl01}
\begin{equation}\label{14}
B^{(1)}_{(2)}=B^{(2)}_{(1)}=
\frac{3Mu_\parallel u_\perp}
{r^3\sqrt{u_4u^4-1}}\left(1-\frac{2M}{r}\right)^{-1/2},
\end{equation}
\begin{equation}\label{15}
B^{(1)}_{(3)}=B^{(3)}_{(1)}=
\frac{3M u_\perp^2 u^4}
{r^3\sqrt{u_4u^4-1}}\left(1-\frac{2M}{r}\right)^{1/2},
\end{equation}
where  $u_\parallel\equiv dr/ds$ and  $u_\perp\equiv rd\varphi/ds $ are the radial and tangential components of the observer's 4-velocity, and  $M$ is Schwarzschild's mass (the standard Schwarzschild coordinates are used).

Let us analyze expressions (\ref{14}) and (\ref{15}) at different velocities of the observer relative to Schwarzschild's mass. Note that the gravitomagnetic components  are nonzero only at $u_\perp\ne 0$.  Then it is easy to see that the components (\ref{14}) and (\ref{15}) significantly depend on the velocity of an observer relative to the Schwarzschild mass. Indeed, at
$ |u_\perp|\ll 1 $, $ |u_\parallel|\ll 1 $,
 the common term $3M/r^3$ in expressions (\ref{14}) and (\ref{15}) is multiplied on the corresponding small values, whereas in the highly relativistic case, for $ |u_\perp|\gg 1 $, this term is multiplied by the large (as compare to 1) values and then
\begin{equation}\label{16}
B^{(1)}_{(2)}=B^{(2)}_{(1)} \sim \frac{3M}{r^3}\gamma,\quad
B^{(1)}_{(3)}=B^{(3)}_{(1)} \sim \frac{3M}{r^3}\gamma^2,
\end{equation}
where $ \gamma $ is the relativistic Lorentz factor as calculated by the particle velocity relative to the Schwarzschild mass \cite{Pl01}.

The acceleration components (\ref{13}) depend, in the case of highly relativistic nonradial motions, on $\gamma $ such that  $ a_{(2)} \sim \gamma $, $ a_{(3)} \sim \gamma^2$. The component $ a_{(1)}$ remains equal to zero at any velocity. The absolute value of the 3-acceleration is proportional to $ \gamma^2 $.

So, according to the MP equations, from the point of view of the observer comoving with a spinning particle in  Schwarzschild's background, the spin-gravity interaction is much greater at the highly relativistic particle's velocity than at the low velocity. This interaction has the clear feature of the spin-orbit interaction. However, it is interesting to estimate the effects of this  interaction for another observer, e.g., which is at rest relative to the Schwarzschild mass.

Another approach to describe the case of a relativistic spinning particle in 
the Schwarzschild field is developed in \cite{Silenko} where in particular the dependence of the spin-gravity interaction on the particle energy is considered.

In the next section we stress that the value of the spinning particle velocity relative to the source of the gravitational field is important in choosing an appropriate supplementary condition.

\section{On adequate supplementary condition for highly relativistic particle motions}

Another example of the spinning particle motions in Schwarzschild's background shows that in the highly relativistic regime some restriction arises on using supplementary condition (\ref{4}). Indeed, for the equatorial motions (with $\theta=\pi/2$ in the standard Schwarzschild
coordinates) of a spinning particle it follows from exact MP equations (\ref{1}) and (\ref{2}) under condition (\ref{4}) that the relationships take place \cite{Pl11}:
$$
P^r=\mu u^r R\left(1-2\varepsilon^2\frac{M}{r}\right),
$$
$$
P^\varphi =\mu u^{\varphi} R\left(1+\varepsilon^2\frac{M}{r}\right),
$$
\begin{equation}\label{17}
P^t=\mu u^t R\left(1-2\varepsilon^2\frac{M}{r}\right),
\end{equation}
where
\begin{equation}\label{18}
R=\left[\left(1-2\varepsilon^2\frac{M}{r}\right)^2-3(u^\varphi)^2\varepsilon^2
Mr\left(2-\varepsilon^2\frac{M}{r}\right)\right]^{-1/2},
\end{equation}
and
\begin{equation}\label{19}
\varepsilon\equiv\frac{|S_0|}{\mu M}
\end{equation}
(for the equatorial motions $P^\theta =0$ and $u^\theta =0$). The constant $\mu$ is the rest mass of the particle and, as a result of the test condition for the spinning particle, it is necessary $\varepsilon\ll 1$ \cite{Wald}. Relations (\ref{18}) and (\ref{19}) illustrate that in the linear 
spin approximation the momentum $P^\lambda$ is parallel to the velocity
$u^\lambda$.

An important result follows from (\ref{17}) and (\ref{18}) due to the terms
with $\varepsilon^2$. Indeed, if 
\begin{equation}\label{20}
|u_\perp|>\frac{\sqrt{r}}{\varepsilon\sqrt{6M}},
\end{equation}
where $u_\perp\equiv ru^\varphi$ is the particle's tangential
velocity, in  (\ref{18}) we have the square root of the negative value. It means that according to  (\ref{17}) and (\ref{18}) under condition (\ref{20})  the expressions for the components of 4-momentum $P^\lambda$ become imaginary for the real mass $\mu$: a result which is not acceptable from the physical point of view. Moreover, for the value $m$, which by (\ref{5}) is equal to $P^\nu u_\nu$, according to (\ref{17}) we have 
\begin{equation}\label{21}
m=\frac{\mu}{R}\left(1-2\varepsilon^2\frac{M}{r} - 3u_\perp^2\varepsilon^2\frac{M}{r} \right).
\end{equation}
It follows from (\ref{21}) that when condition (\ref{20}) is satisfied and $R$ is imaginary, the value $m$ becomes imaginary as well. However, then according to  (\ref{5}) in the left-hand part of this relationship we have the
pure imaginary value and in the right-hand part there is the sum of the imaginary value $mu^\nu$ and the second real term (the value $u_\lambda DS^{\nu\lambda}/ds$ is real because by definition and its physical meaning, the real value is $S^{\nu\lambda}$). Therefore, we conclude that in the case of condition (\ref{4}), when  relationship  (\ref{20}) takes place, from the exact Mathisson-Papapetrou equations follows the result which is not satisfactory from the physical point of view.

The important feature of relationship (\ref{20}) is that the orbital velocity
value must be highly relativistic for all values of the radial coordinate beyond the horizon surface, because here $\varepsilon\ll 1$. In particular, 
if $r$ is not much greater than $M$, the velocity value of
the right-hand side of Eqn. (\ref{20}) corresponds to the particle's
highly relativistic Lorentz $\gamma$-factor of order $1/\varepsilon$.
It means that supplementary condition (\ref{4}) cannot be used for the particle's velocity which is very close to the velocity of light. In this context we stress that according to relationship (\ref{20})  an unexpected conclusion which is discussed in \cite{Hojman:2013} becomes clear. Namely,
in \cite{Hojman:2013} it is shown that the MP equations at condition (\ref{4}) have solutions which allow acceleration of a spinning particle to the superluminal velocity in Schwarzschild's background. Note that  under  condition (\ref{3}) the corresponding result is not allowed by the MP equations.

Thus, in the linear spin approximation one can use the supplementary condition for the MP equations in form (\ref{3}) or (\ref{4}). However, when the nonlinear spin terms are taken into account and the particle velocity is so high as in (\ref{20}) then the acceptable condition is only (\ref{3}).

\section{Effects of  highly relativistic spin-gravity coupling on particle trajectories}

So, in Secs. III and IV above we considered  two different situations which demonstrate the specific features of a spinning particle motion in the highly relativistic regime according to the MP equations. Now we briefly analyze how this regime reflects in the particle's trajectory.

\subsection{Linear  spin approximation}

Here we discuss results following from the MP equations in the linear spin approximation for the highly relativistic motions in the concrete metrics which is common for conditions (\ref{3}) and (\ref{4}).

First, we point out the circular highly relativistic orbits of a spinning particle which are allowed by the MP equations in the equatorial plane of the Schwarzschild source with the spin orthogonal to this plane. These orbits exists in the small neighborhood of the radial coordinate value $r=1.5 r_g(1+\delta)$, where $r_g$ is the horizon radius, $|\delta|\ll 1$ and it is allowed $\delta$ to be positive, negative or zero \cite{Pl05}. [For comparison, the highly relativistic circular orbits of a spinless particle are allowed by the geodesic equation only for $\delta>0$ \cite{Chandra}.] For realization of these orbits the spinning particle must possess the orbital velocity which corresponds to the Lorentz $\gamma$-factor of order $1/\sqrt{\varepsilon}$, and this velocity is highly relativistic in the sense that
$\gamma^2\gg 1$, because of  $\varepsilon\ll 1$. The similar highly relativistic circular orbits exist in the equatorial plane of Kerr's source for
the radial coordinate values $r= r_{ph}^{(-)}(1+\delta)$,  where $r_{ph}^{(-)}$ is the Boyer-Lindquist radial coordinate of the counterrotating circular photon orbits \cite{Pl10}. All these circular orbits are caused by the strong attractive action of the spin-gravity coupling in addition to the geodesic attraction. Moreover, if the initial value of the spinning particle velocity  slightly differs from the corresponding value for these highly relativistic circular orbits (for example, due to some nonzero value of the
particle's radial velocity), the corresponding noncircular orbits illustrate the
situations when for the short time of the spinning particle motion relative to  Schwarzschild's or Kerr's mass the space deviation of this particle trajectory from the geodesic one becomes significant \cite{Pl12, Pl13}.

\subsection{Nonlinear effects}

According to the exact MP equations at condition (\ref{3}) the nonlinear spin terms become important in many situations with a fast moving spinning
particle in Schwarzschild's or Kerr's metric \cite{Pl12, Pl13}. For example, in contrast to the cases of the linear spin approximation, when the highly relativistic circular orbits of a spinning particle exist in the narrow space regions only, due to the nonlinear terms the corresponding domains are much wider. 

In particular, by the exact MP equations, the highly relativistic equatorial circular orbits of a spinning particle in Kerr's background are allowed not only for $r= r_{ph}^{(-)}(1+\delta)$ with $|\delta|\ll 1$, but for any $r= r_{ph}^{(-)}$, up to $r\gg r_{ph}^{(-)}$. However, in the last case the necessary value of the Lorentz $\gamma$-factor is much greater than, for example, in the case for $r=r_{ph}^{(-)}$. All these orbits, similarly to the corresponding orbits following from the MP equations in the linear spin approximation, are caused by the strong attractive action of the spin-gravity coupling.

Due to the strong repulsive action of the spin-gravity coupling on the particle, there are the highly relativistic circular orbits of a spinning particle in the space region with $r< r_{ph}^{(+)}$ \cite{Pl13} in the equatorial plane of the Kerr background. In addition, the same repulsive force gives the significantly nongeodesic circular orbits of a spinning particle beyond the equatorial planes in the Schwarzschild and Kerr backgrounds  
\cite{Pl82}.

\subsection{Numerical estimates}

Do some particles in cosmic rays possess the sufficiently high $\gamma$-factor for motions on the highly relativistic circular orbits, or on some their fragments, in the gravitational field of the Schwarzschild or Kerr black hole which are considered above? Yes, they do.
By the numerical estimates  for an electron in the gravitational field of a black hole with three of the Sun's mass the value $|\varepsilon_0$ is equal $4\times 10^{-17}$. Then the necessary value of the $\gamma$-factor for the realization of some highly relativistic circular orbits by the electron near this black hole is of order $10^8$. This $\gamma$-factor corresponds to the energy of the electron free motion of order $10^{14}$ eV. Analogously, for a proton in the field of such a black hole the corresponding energy is of order  $10^{18}$ eV. For the massive black hole those values are greater: for example, if $M$ is equal to $10^6$ of the Sun's mass the corresponding value of the energy for an electron is of order $10^{17}$ eV and for a proton it is $10^{21}$ eV.
Naturally, far from the black hole these values are greater because the necessary $\gamma$-factor is proportional to $\sqrt{r}$.

Note that for a neutrino near the black hole with three of the Sun's mass the necessary values of its $\gamma$-factor for motions on the highly relativistic circular orbits correspond to the neutrino's energy of the free motion of order $10^5$ eV. If the black hole's mass is of order $10^6$ of the Sun's mass, the corresponding value is of order $10^8$ eV.

Can the highly relativistic spin-gravity effects be registered by the observation of the electromagnetic synchrotron radiation from some black holes? Perhaps, however, it is difficult to differ the situation when the circular orbits of a spinning charged particle and its synchrotron radiation are caused by the magnetic field or the spin-gravity coupling. The detailed analysis of the observational data is necessary.

\section{On quantum equation \\ for highly relativistic fermions \\ in a gravitational field}

Naturally, a more appropriate description of the highly relativistic spin-gravity coupling for electrons and other fermions cannot be restricted by the MP equations and it is necessary to analyze the corresponding quantum equations. As an example, in this context we point out paper \cite{Pl01} where the solution of the Dirac equation in a Schwarzschild field which describes the quantum state corresponding to the classical highly relativistic orbit with $r=1.5 r_g$ is considered. However, in general the situation with the Dirac equation is not so simple.

Indeed, it is shown in many papers that in the linear spin approximation the MP equations follow from the general relativistic Dirac equation as some classical approximation. Here we draw attention to the fact
that the exact MP equations (i. e., their nonlinear  spin terms) cannot be obtained from this Dirac equation in principle. Why? To answer this question 
we recall that the main step in obtaining the general relativistic Dirac equation in the curved spacetime consists in introduction the notion of the parallel transport for spinors as a generalization of this notion for tensors, whereas according to the MP equations the spin of a test particle is transported by Fermi:
\begin{equation}\label{22}
\frac{Ds^\mu}{ds}=u^\mu \frac{Du_\nu}{ds}s^\nu.
\end{equation}
 It follows from (\ref{22}) that only in the linear spin approximation
the Fermi transport coincides with the parallel transport. Therefore,  to satisfy the principle
of correspondence between the Dirac equation and the exact MP equations, at first sight,  it is necessary simply to introduce and use the Fermi transport for spinors in some corrected Dirac equation. However, it is impossible without the Lorentz invariance violation.  In this context we note that  many papers are devoted to the violation of Lorentz invariance from different points of view, for example, in the context of the Standard-Model Extension  \cite{Kostel, Bonder}and, probably,  the key words {\it Lorentz-violating spinor} first appeared only recently \cite{Bonder}. One can hope that just  in the framework of this approach the necessary corrected Dirac equation will be obtained. Some preliminary estimates show that the correspondent
equation must be nonlinear in $\psi$-function \cite{PlArx}. (Note, from another position and in another context, some modified Dirac equation with Lorentz invariance violation was proposed in  \cite{Krug}.) In any case, further investigations in this direction are necessary.

\section{Conclusions}

Concerning the question from the first paragraph in the Introduction of this paper we conclude: there is the significant difference in the motions of a spinning particle with low and very high velocities relative to the source of a gravitational field. This result follows from the solutions of the MP equations in Schwarzschild's and Kerr's metric, both under conditions (\ref{3}) and (\ref{4}), both in the linear and nonlinear spin consideration. The force which deviates the motion of a spinning particle from the geodesic free fall trajectory is proportional to the second power of the relativistic Lorentz $\gamma$-factor, by the estimation of an observer comoving with this particle.

It is important  that the results presented in Sec. IV indicate the limit of validity of condition (\ref{4}): for the high orbital particle velocity in sense (\ref{20}) this condition is not adequate from the physical point of view.

Probably, the MP equations can play an important heuristic role in finding some corrections to the general relativistic Dirac equation for a more adequate description of highly relativistic fermions in a gravitational field.

\end{document}